\documentclass[twocolumn,showpacs,preprintnumbers,epsfig,prl,aps]{revtex4}
\usepackage{amssymb}

\usepackage{graphicx}
\usepackage{dcolumn}
\usepackage{bm}

\begin{document}

\preprint{}
\title{On the occurrence of Berezinskii-Kosterlitz-Thouless behavior in highly
anisotropic cuprate superconductors}
\author{T. Schneider}
\affiliation{Physikinstitut, University of Zurich, Winterthurerstrasse 190, 8057 Zurich,
Switzerland. }
\date{\today }

\begin{abstract}
The conflicting observations in the highly anisotropic
Bi$_{2}$Sr$_{2}$CaCu$_{2}$O$_{8+\delta }$, evidence for BKT behavior
emerging from magnetization data and smeared 3D-xy behavior,
stemming form the temperature dependence of the magnetic in-plane
penetration depth are traced back to the rather small ratio, $\xi
_{c}^{+}/\xi _{c}^{-}=\xi _{c0}^{+}/\xi _{c0}^{-}\simeq 0.45$,
between the $c$-axis correlation length probed above ($\xi
_{c}^{+}$) and below ($\xi _{c}^{-}$) $T_{c}$ and the comparatively
large anisotropy. The latter leads to critical amplitudes $\xi
_{c0}^{\pm }$ which are substantially smaller than the distance
between two CuO$_{2}$ double layers. In combination with $\xi
_{c}^{+}/\xi _{c}^{-}\simeq 0.45$ and in contrast to the situation
below $T_{c}$ the c-axis correlation length $\xi _{c}^{+}$ exceeds
the distance between two CuO$_{2}$ double layers very close to
$T_{c} $ only. Below this narrow temperature regime where 3D-xy
fluctuations dominate, there is then an extended temperature regime
where the units with two CuO$_{2}$ double layers are nearly
uncoupled so that 2D thermal fluctuations dominate and BKT features
are observable.
\end{abstract}

 \maketitle

Since the pioneering work of Berezinskii\cite{bere}, Kosterlitz and
Thouless\cite{kt} (BKT) on the BKT transition in the two-dimensional
(2D) XY model, much efforts have been devoted to observe the
universal behavior characteristic of the KT transition, as the
universal jump of the superfluid density\cite{nelson}, measured in
$^{4}$He superfluid films, or the non-linear I - V characteristic,
observed in thin films of conventional
superconductors\cite{minnhagen,tsbook}. Signatures of BKT physics
can be expected also in layered superconductors with weak
inter-layer coupling. Potential candidates are underdoped cuprate
superconductors where the anisotropy increases with reduced
transition temperature $T_{c}$\cite{tsphysb}. Recent studies of the
I-V characteristic\cite{vadla}, the frequency dependent
conductivity\cite{corson}, the Nernst signal\cite{wang}, the
magnetization\cite{li,wang2}, and of the resistance\cite{matthey}
have been interpreted as signatures of BKT behavior. On the other
hand, several experiments
\cite{jacobs,pana1,pana2,osborn,hosseini,broun,liang,zuev,ruf}
failed to observe any trace of the universal jump in the superfluid
density around $T_{c}$. Indeed, a systematic finite-size scaling
analysis of in-plane penetration depth data taken on films and
single crystals of the highly anisotropic
Bi$_{2}$Sr$_{2}$CaCu$_{2}$O$_{8+\delta }$ reveal a smeared
transition, consistent with an inhomogeneity induced finite-size
effect\cite{tscastro}. Furthermore, there is evidence that in small
thin-film samples on insulating substrates, edge effects modify the
vortex-vortex interaction making it short-range, unlike the
logarithmic long-range interaction needed for the BKT transition.
This appears to make the BKT transition impossible in thin films of
any size if they are supported by a non-superconducting
substrate\cite{kogan}.

\begin{figure}[tbp]
\centering
\includegraphics[angle=0,width=8.6cm]{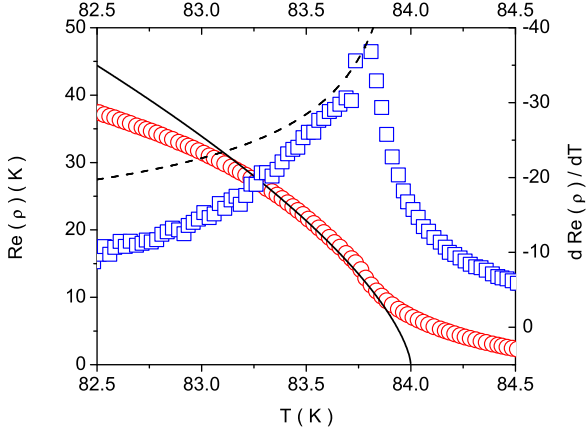}
\caption{Re$(\rho )\propto 1/\lambda _{ab}^{2}$ and $d$Re$(\rho
)/dT$ \textit{vs}. $T$ for the
Bi$_{2}$Sr$_{2}$CaCu$_{2}$O$_{8+\delta }$ film B, $616$ \AA\ thick,
derived from Osborn \textit{et al}.\protect\cite{osborn} The solid
and dashed lines indicate the leading critical behavior of a
homogeneous bulk system according to Eq. \ref{eq1}, where Re$(\rho
)=650\left\vert t\right\vert ^{2/3}$ with $T_{c}=84$ K. }
\label{fig1}
\end{figure}

In this work we attempt to unravel the conflicting observations on
the highly anisotropic Bi$_{2}$Sr$_{2}$CaCu$_{2}$O$_{8+\delta }$:
evidence for BKT behavior emerging from the magnetization data of Li
\textit{et al. }\cite{li} and evidence for smeared 3D-xy behavior,
stemming form the temperature dependence of the magnetic in-plane
penetration depth \cite{jacobs,osborn,tscastro}. Although
Bi$_{2}$Sr$_{2}$CaCu$_{2}$O$_{8+\delta }$ is highly anisotropic
\cite{kawamata}, resistivity measurements \cite{watanabe} uncover
clearly the three dimensional nature of the transition in this
extreme type II superconductor. Accordingly, sufficiently close to
$T_{c}$ homogeneous samples are expected to exhibit 3D-xy critical
behavior. To explore how close this should be, we consider the
magnetic in-plane penetration data shown in Fig. \ref{fig1}, derived
from the complex conductivity measurements of Osborn \textit{et al}.
\cite{osborn} on epitaxially grown
Bi$_{2}$Sr$_{2}$CaCu$_{2}$O$_{8+\delta }$ films, using a two-coil
inductive technique at zero applied. For comparison we included the
leading 3D-xy behavior given by the universal relation \cite
{tsbook, tscastro,parks,peliasetto}

\begin{equation}
\frac{1}{\lambda _{ab}^{2}\left( T\right) }=\frac{16\pi
^{3}k_{B}T}{\Phi _{0}^{2}\xi _{c}^{-}\left( T\right) },~ \xi
_{c}^{-}=\xi _{c0}^{-}\left\vert t\right\vert ^{-\nu },  \label{eq1}
\end{equation}
with
\begin{equation}
t=T/T_{c}-1,~ \nu \simeq 2/3.  \label{eq2}
\end{equation}

Apparently, there is a rounded transition pointing to a finite size
effect, preventing the $c$-axis correlation length $\xi _{c}^{-}$ to
grow beyond the limiting length $L_{c}$, set by inhomogeneities.
Indeed the extreme in $d$Re$\left( \rho \right) /dT$ exhibits an
inflection point at $T=T_{p_{c}}\simeq 83.77$ K where $\xi _{c}^{-}$
attains the limiting length $L_{c}$. As shown previously
\cite{tscastro}, Eq. (\ref{eq1}) yields together with $T_{c}=84$ K,
$T_{p_{c}}\simeq 83.77$ K, Re$\left( \rho \left( T_{c}\right)
\right) /$Re$\left( \rho \left( T=0\right) \right) =0.011$,
Re$\left( \rho \left( T=0\right) \right) =0.6$ K, Re$\left( \rho
\left( T_{p_{c}}\right) \right) =13.66$ K, $\lambda _{ab}\left(
0\right) =265$ nm, and $L_{c}=\xi _{c}^{-}\left( T_{p_{c}}\right) $
the estimates
\begin{equation}
\xi _{c0}^{-}\simeq 2{\AA},~L_{c}\simeq 93{\AA}, \label{eq3}
\end{equation}
for the critical amplitude of the $c$- axis correlation length below
$T_{c}$ and the limiting length $L_{c}$ set by inhomogeneities. The
rather small value of $\xi _{c0}^{-}$,  reflecting the high
anisotropy, $\gamma =\xi _{ab0}^{-}/\xi _{c0}^{-}$, was confirmed in
other Bi$_{2}$Sr$_{2}$CaCu$_{2}$O$_{8+\delta }$ films and single
crystals in terms of a detailed finite size scaling analysis
\cite{tscastro}.

\begin{figure}[tbp]
\centering
\includegraphics[angle=0,width=8.6cm]{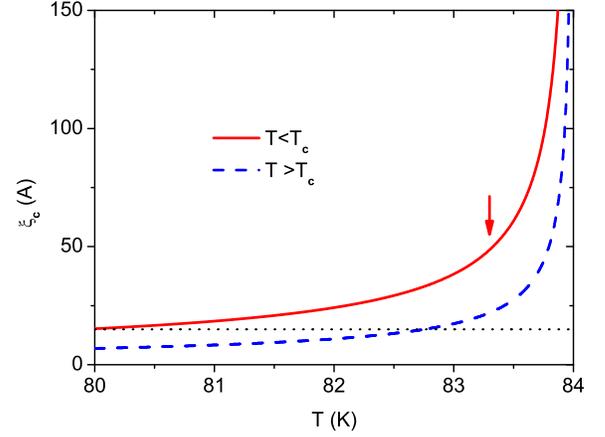}
\caption{$\xi _{c}^{-}$ and $\xi _{c}^{+}$ \textit{vs}. $T$ for $\xi
_{c0}^{-}=2$ \AA , $\xi _{c0}^{+}=0.9$ \AA\ and $T_{c}=84$ K, where
$\xi _{c}^{\pm }=\xi _{c0}^{\pm }\left\vert t\right\vert ^{-2/3}$.
The dotted line marks $s=15$\AA , the distance between two CuO$_{2}$
double layers. The arrow marks $T=83.25$ K where in Fig. \ref{fig1}
consistency with 3D-xy critical behavior sets in.} \label{fig2}
\end{figure}

Although the $c$-axis correlation length $\xi _{c}^{-}$ increases by
approaching $T_{c}$, the occurrence of 3D thermal fluctuations
requires that in Bi$_{2}$Sr$_{2}$CaCu$_{2}$O$_{8+\delta }$ it
exceeds $s=15$ \AA , the distance between two CuO$_{2}$ double
layers. In Fig. \ref{fig2} we depicted the temperature dependence of
$\xi _{c}^{-}$. Even though it exceeds $15$ \AA\ around $80$ K , a
glance to Fig. \ref{fig1} shows that consistency with 3D-xy behavior
is achieved much closer to $T_{c}$ only, namely above $T\approx
83.25$ K. Accordingly, the occurrence of 3D-xy behavior requires
$\xi _{c}^{-}$ to exceed the distance between two CuO$_{2}$ double
layers by at least a factor of three. Above $T_{c}$ the situation is
even worse because the ratio between the correlation lengths above
($\xi _{c}^{+}$) and below ($\xi _{c}^{-}$) $T_{c}$ is a universal
quantity and in the 3D-xy universality class given by
\begin{equation}
\xi _{c0}^{+}/\xi _{c0}^{-}\simeq 0.45.  \label{eq4}
\end{equation}
This ratio implies a substantial shrinkage of the 3D-xy fluctuation
dominated regime above $T_{c}$. To illustrate this point we included
in Fig. \ref{fig2} the temperature dependence of $\xi _{c}^{+}$.
Taking again $\xi _{c}^{+}\gtrsim 3s=45$ \AA\ to locate the regime
where 3D-xy fluctuations dominate we obtain roughly $T\gtrsim 83.8$
K which is rather close to the $T_{c}\simeq 84$ K of the fictitious
homogeneous system.

\begin{figure}[tbp]
\centering
\includegraphics[angle=0,width=8.6cm]{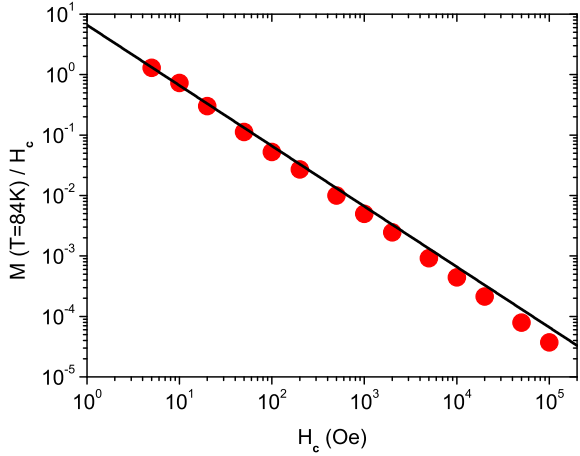}
\caption{$M/H$ \textit{vs}. $H$ for
Bi$_{2}$Sr$_{2}$CaCu$_{2}$O$_{8+\delta }$ at $T=84$  K derived from
the magnetization data of Li \textit{et al.} \protect\cite{li} with
$H$ applied along the $c$-axis. The solid line is $M(T=84K)/H=6.6/H
$ corresponding to Eq. (\ref{eq7}).} \label{fig3}
\end{figure}

Given the reduced 3D-xy critical regime above $T_{c}$ the system
corresponds to a stack of nearly independent units with two
CuO$_{2}$ double layers as long as $\xi _{c}^{+}\lesssim 3s$. In
this intermediate regime, 2D and in particular BKT features are then
expected. To check this expectation quantitatively we consider the
3D-xy scaling expression for the susceptibility in the limit
$T\gtrsim T_{c}$ and $H_{c}\rightarrow 0$
\cite{tsbook,tseuro,tsjh,tsjh2}
\begin{equation}
\frac{m}{H_{c}}=-\frac{Q^{+}C_{3,0}^{+}k_{B}T\left( \xi
_{ab}^{+}\right) ^{2}}{\Phi _{0}^{2}\xi _{c}^{+}},  \label{eq5}
\end{equation}
where $m=M/V$ is the magnetization per unit volume and
$Q^{+}C_{3,0}^{+}\simeq 0.9$ a universal number. In type II
superconductors, exposed to a magnetic field $H_{i}$ in direction
$i$, there is also the magnetic field induced limiting length
$L_{H_{i}}=\sqrt{\Phi _{0}/\left( aH_{i}\right) }$ with $a\simeq
3.12$ \cite{bled}, related to the average distance between vortex
lines \cite{parks,bled}. As the magnetic field increases, the
density of vortex lines becomes greater, but this cannot continue
indefinitely, the limit is roughly set on the proximity of vortex
lines by the overlapping of their cores. Due to these limiting
length the phase transition is rounded and occurs smoothly. Indeed,
approaching $T_{c}$ from above the correlation lengths combination
$\xi _{i}^{+}\xi _{j}^{+}$ increases but is bounded by
$L_{H_{k}}^{2}=\Phi _{0}/\left( aH_{k}\right) $ where $i\neq j\neq
k$. In this context it is important to recognize that the
confinement effect of the magnetic field in direction i on
fluctuations within a region $L_{H_{i}}$ acts only in the plane
perpendicular to the field. Therewith Eq. (\ref{eq5}) reduces for
$T\simeq T_{c}$ and $H_{c}\rightarrow 0$ to
\begin{equation}
\frac{m}{H_{c}}=-\frac{Q^{+}C_{3,0}^{+}k_{B}T\gamma }{\Phi
_{0}^{3/2}a^{1/2}H_{c}^{1/2}},  \label{eq6}
\end{equation}
because $\xi _{ab}^{+}/\xi _{c}^{+}=\gamma $ and the growth of $\xi
_{ab}^{+} $ is limited by $\xi _{ab}^{+}=L_{H_{c}}\simeq \sqrt{\Phi
_{0}/\left( aH_{c}\right) }$. On the other hand, in the strict 2D
case $\xi _{c}^{+}$ cannot grow beyond the thickness $d_{s}$ of the
system. Consequently, as $-m/H_{c}$ and $\xi _{ab}^{+}$ initially
increase with reduced temperature in a fixed field they then
saturate due to the magnetic field induced finite size effect. In
this case Eq. (\ref{eq5}) reduces to
\begin{equation}
\frac{m}{H_{c}}=-\frac{Q^{+}C_{3,0}^{+}k_{B}T}{\Phi
_{0}d_{s}aH_{c}}. \label{eq7}
\end{equation}
In Fig. \ref{fig3} we show $-M/H$ \textit{vs}. $H$ for
Bi$_{2}$Sr$_{2}$CaCu$_{2}$O$_{8+\delta }$ close to $T_{c}$ at $T=84$
K derived from the magnetization data of Li \textit{et al.}
\cite{li} , with $H$ applied along the $c$-axis. The comparison with
the characteristic 2D-behavior (\ref{eq7}) reveals remarkable
agreement, extending over nearly two decades of the applied magnetic
field. Noting that down to the lowest applied magnetic field,
$H_{c}=5$ Oe, there is no sign of leveling off, arising from an
inhomogeneity induced finite size effect, it follows that the length
of the homogenous regions in the $ab$-plane, $L_{ab}$, exceeds the
attained magnetic field induced limiting length
$L_{H_{c}}=\sqrt{\Phi _{0}/\left( aH_{c}\right) }=1.14$ $10^{-4}$
cm. This behavior does not confirm the occurrence of intermediate 2D
critical behavior in sufficiently anisotropic systems above $T_{c}$
only. In addition, it uncovers that in the $ab$-plane excellent
homogeneity can be achieved and an applied magnetic field leads to
the outlined finite size effect.

\begin{figure}[tbp]
\centering
\includegraphics[angle=0,width=8.6cm]{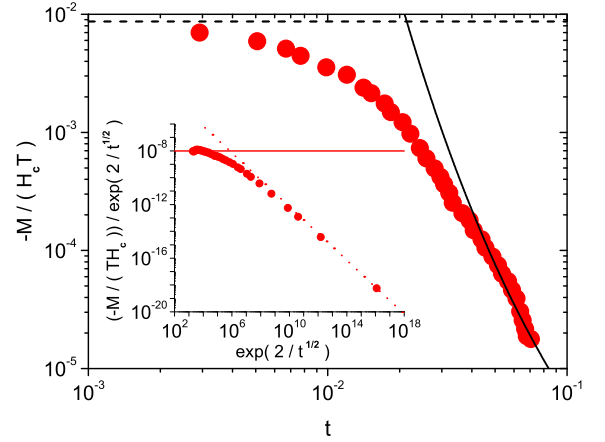}
\caption{$-M/\left( TH_{c}\right) $ \textit{vs}. $t=T/T_{c}-1$ for
Bi$_{2}$Sr$_{2}$CaCu$_{2}$O$_{8+\delta }$ at $H_{c}=10$ Oe with
$T_{c}=84$ K derived from the magnetization data of Li \textit{et
al.} \protect\cite{li}. The solid line is $-M/\left( TH_{c}\right)
=10^{-8}\exp (2/t^{1/2})$ corresponding to Eq. (\ref{eq8}) with $\xi
_{ab}^{+}\propto \exp (bt^{-1/2})$ with $b=1$ and the dashed one
$-M/\left( TH_{c}\right) =8.7$ $10^{-3}$. The inset shows the
finite-size scaling function in terms of $\left( -M/\left(
TH_{c}\right) \right) /\exp (2bt^{-1/2})$ \textit{vs}. $\exp
(2bt^{-1/2})$ with $b=1$. The dotted line is $\left( -M/\left(
TH_{c}\right) \right) /\exp (2bt^{-1/2})=0.007/$ $\exp
(2bt^{-1/2})$, indicating the limiting magnetic field induced
finite-size behavior and the solid one is $\left( -M/\left(
TH_{c}\right) \right) /\exp (2bt^{-1/2})=10^{-8}$ marking the BKT
limit.} \label{fig4}
\end{figure}

To substantiate the occurrence of the magnetic field induced finite
size effect further, we consider the temperature dependence of the
magnetization at fixed magnetic field. When $\xi _{c}^{+}<3s$ and
$L_{H_{c}}>>\xi _{ab}^{+} $ Eq. (\ref{eq5}) reduces to
\begin{equation}
\frac{m}{TH_{c}}=-\frac{Q^{+}C_{3,0}^{+}k_{B}\left( \xi
_{ab}^{+}\right) ^{2}}{\Phi _{0}^{2}d_{s}}.  \label{eq8}
\end{equation}
In this regime the in-plane correlation length $\xi _{ab}^{+}$ is
expected to exhibit the characteristic BKT-behavior in the reduced
temperature $t=T/T_{c}-1$, namely $\xi _{ab}^{+}=a\exp
(bt^{-1/2})$\cite{kt}, with $b\approx 1$ a non-universal constant
and $a$ related to the vortex core diameter. However, as $t$
decreases and with that $\xi _{ab}^{+}$ approaches the limiting
length $L_{H_{c}}=\sqrt{\Phi _{0}/\left( aH_{c}\right) }$ a finite
size effect sets in and $\xi _{ab}^{+}$ saturates in the limit
$L_{H_{c}}<<\xi _{ab}^{+}$ to $\xi _{ab}^{+}=L_{H_{c}}$. To describe
the resulting crossover we introduce the finite-size scaling
function $S$ \cite{cardy} in terms of $\left( \widetilde{\xi
_{ab}^{+}}\right) ^{2}=\left( \xi _{ab}^{+}\right) ^{2}S\left(
H_{c}\left( \xi _{ab}^{+}\right) ^{2}/\Phi _{0}\right) =\left( \xi
_{ab}^{+}\right) ^{2}S\left( \left( \xi _{ab}^{+}\right) ^{2}/\left(
aL_{H_{c}}^{2}\right) \right) $ where $S\left( \left( \xi
_{ab}^{+}\right) ^{2}/\left( aL_{H_{c}}^{2}\right) \right)
\rightarrow 1$ for $L_{H_{c}}>>\xi _{ab}^{+}$, $S\left( \left( \xi
_{ab}^{+}\right) ^{2}/\left( aL_{H_{c}}^{2}\right) \right)
\rightarrow saL_{H_{c}}^{2}/\left( \xi _{ab}^{+}\right) ^{2}$ for
$L_{H_{c}}<<\xi _{ab}^{+}$, and substitute $\widetilde{\xi
_{ab}^{+}}$ in Eq. (\ref{eq8}) instead of  $\xi _{ab}^{+}$. In Fig.
\ref{fig4} we depicted $-M/\left( TH_{c}\right) $ \textit{vs}.
$t=T/T_{c}-1$ at $H_{c}=10$ Oe and the inset shows the finite-size
scaling function in terms of $\left( -M/\left( TH_{c}\right) \right)
/\exp (2bt^{-1/2})$ \textit{vs}. $\exp (2bt^{-1/2})$. It is seen
that by approaching the transition temperature $-M/\left(
TH_{c}\right) $ increases but saturates to $-M/\left(
T_{c}H_{c}\right) =8.7$ $10^{-3}$, in agreement with the behavior
shown in Fig. \ref{fig3}. Sufficiently away from $T_{c}$ a crossover
to the BKT behavior, requiring $L_{H_{c}}>>\xi _{ab}^{+}$ and
indicated by the solid line, can be anticipated. As the crossover
extends over a rather extended temperature regime agreement with the
leading BKT behavior is limited and the 3D-xy critical regime is not
accessible. Nevertheless, the finite-size scaling function reveals
the flow to BKT behavior is attained. To provide a consistency test
we use $M(T=84K)/H_{c}=6.6/H_{c}$ (Fig. \ref{fig3}) and $d_{s}=15$
\AA\ to estimate $a$ from $\left( -M/\left( TH_{c}\right) \right)
/\exp (2bt^{-1/2})=10^{-8}$ (\ref{fig4}) with the aid of Eqs.
(\ref{eq7}) and (\ref{eq8}). The result is $a\simeq 9.2$ \AA, in
reasonable agreement with the estimate of Li \textit{et al}.
\cite{li}.

We have seen that the conflicting observations in the highly
anisotropic Bi$_{2}$Sr$_{2}$CaCu$_{2}$O$_{8+\delta }$, evidence for
BKT behavior emerging from the magnetization data of Li \textit{et
al.}\cite{li} and smeared 3D-xy behavior, stemming form the
temperature dependence of the magnetic in-plane penetration
depth\cite{jacobs,osborn,tscastro} are a consequence of the rather
small ratio, $\xi _{c}^{+}/\xi _{c}^{-}=\xi _{c0}^{+}/\xi
_{c0}^{-}\simeq 0.45$, between the $c$-axis correlation length
probed above ($\xi _{c}^{+}$) and below ($\xi _{c}^{-}$) $T_{c}$ and
the comparatively large anisotropy. The latter leads to critical
amplitudes $\xi _{c0}^{\pm }$ which are substantially smaller than
the distance between two CuO$_{2}$ double layers. In combination
with $\xi _{c}^{+}/\xi _{c}^{-}\simeq 0.45$ and in contrast to the
situation below $T_{c}$ the c-axis correlation length $\xi _{c}^{+}$
exceeds the distance between two CuO$_{2}$ double layers very close
to $T_{c}$ only. Below this narrow temperature regime where 3D-xy
fluctuations dominate, there is then an extended temperature regime
left where the  units with two CuO$_{2}$ double layers are nearly
uncoupled so that 2D thermal fluctuations dominate and BKT features
are observable. As this behavior relies on the universal ratio $\xi
_{c}^{+}/\xi _{c}^{-}\simeq 0.45$ and pronounced anisotropy it is
not an artefact of Bi$_{2}$Sr$_{2}$CaCu$_{2}$O$_{8+\delta }$, but a
generic feature of sufficiently anisotropic cuprate superconductors.
Examples include underdoped La$_{2-x}$Sr$_{x}$CuO$_{4}$ and
YBa$_{2}$Cu$_{3}$O$_{7-\delta }$ where the temperature dependence of
the in-plane superfluid density does not reveal any trace of the
universal jump
\cite{jacobs,pana1,pana2,osborn,hosseini,broun,liang,zuev,ruf}.

\end{document}